# Implementation of encryption on telemedicine

*Ulkar Ahmadova, Laman Mammadova, Behnam Kiani Kalejahi*
School of Science and Engineering, Khazar University, Baku, Azerbaijan,

*1. Abstract*
In the era of technologies data security is one of the most important things that both individuals and companies need.
Information plays a huge role in our everyday life and keeping it safe should be our number one priority. Nowadays most of the information is been transferred via the internet. One of the ways to use it is telemedicine.With the help of telemedicine, people can have an appointment at the doctors without losing their time or money. All of the information about one's health is transferred through the internet but is it that safe? What techniques are used to provide the safety of our confidential information?
To guarantee that the information is not changed or that in case it will be stolen no one can still have access to it. In this paper, we will talk about how to keep patient's data secured from unauthorized use.

*2. Introduction*:
Cryptography- is a science of protecting the information from unauthorized use through converting plain text into an inconsequential string. It is one of the most important techniques in data security. Assume that you sent a critical email to your co-worker that has confidential information. In case of an attacker stealing it, your company will be damaged. You can avoid this situation by translating the email into a "secret language" that only he can understand, so when the attacker will get the information, he will see just a string of characters with no meaning. There are 3 main components of information security:
Confidentiality – a set of rules to prevent loss of unauthorized data.
Integrity – ensures that the data is not altered.
Availability – the characteristics of a system that ensures that only authorized users have access to the information.

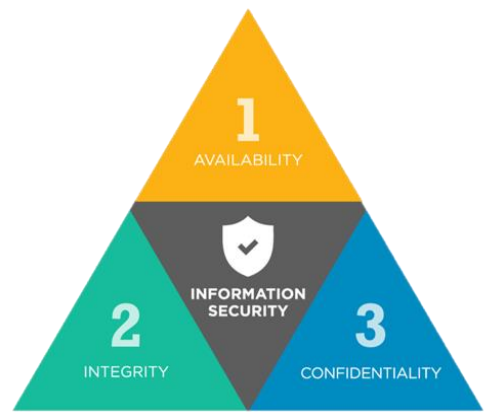

*Fig 1: CIA Triad*

**Terms used in Cryptography**:
Encryption – is the process of transforming plaintext into cipher text.
Decryption – is the process of transforming cipher text back into plaintext.
Key – is a random value that is applied with an algorithm to unencrypted text and converts it into encrypted text.
Plaintext – is a raw and unchanged information. Example: "Hello, World!".
Cipher text – is a string that was created as a result of implementing an encryption key on a plaintext. Example: "0be47nzu0d8d".

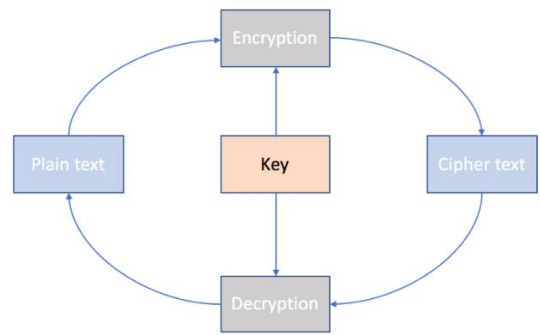

*Fig 2: Cryptography processes*

Encoder – a person who converts plaintext into cipher text.

Decoder – a person who receives cipher text and converts it back into plaintext.

### 3. Symmetric encryption

Symmetric encryption is the type of encryption where the same encryption key is used to both encrypt and decrypt information. Until the 1970s, it was the only type of encryption and even now, it is still the most used one. There are 2 types of cipher – stream cipher and block cipher. Stream cipher encrypts one character at a time. Block cipher first breaks the plaintext into blocks with a fixed number of bits or characters.In case of an infinite or continuous number of bits, it is better to use stream cipher, in case of plaintext with defined length – block cipher.

First, encoder encrypts the plaintext using an algorithm and encryption key. He sends the encrypted text (also called cipher text) and the same encryption key to the decoder. The last one decrypts the cipher text using the key to get the plaintext. Most used symmetric algorithms are DES, 3DES, AES and Blowfish.

**3.1. DES (Data Encryption Standard)** – was developed by IBM and the US government in 1974 but officially published in 1976 by National Institute of Standard and Technology (NIST), being the first standardized encryption system. Besides, it was the first encryption that was approved by National Bureau of Standards (NBS).

DES algorithm divides the data into blocks by 64 bits and encrypts it using a key with length 56 bits. The 64-bit block is halved and makes a 32-bit block that after is expanded into a 48-bit key by expansion permutation (duplicating half of the bits). The process repeats 16 times, so we say that it has 16 rounds. The output, the cipher text, also consists of 64 bits. Its efficiency is slow. This algorithm's performance is better on hardware rather than on software. DES security issue is that it is vulnerable to Brute-force attacks.

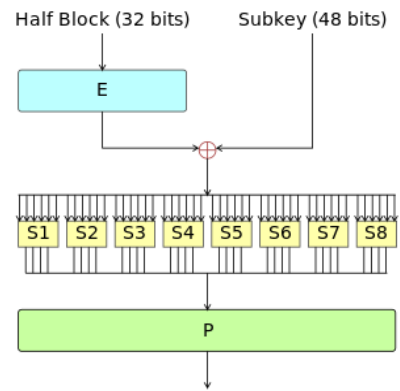

*Fig 3: The Feistel function of DES*

**3.2. 3DES (Triple Data Encryption Standard)** – was developed by IBM in 1998. 3DES divides the data into blocks by 64 bits. The working principle is the same as in DES the main difference is between the sizes of these algorithms. It implements 3 different keys on the block whereas DES implements only one. The key size of DES is 56*3= 168 bits. The number of rounds also triples that gives us 48 rounds. 3DES's effectiveness is also slow, especially on software.

**3.3. Advanced Encryption Standard (AES)** – started to be been developed in 1997 and was published in 2001 by National Institute of Standard and Technology (NIST). Its main purpose was replacing DES after some vulnerabilities were explored. This algorithm is based on Rijndael cipher. AES works with 128-bit block size. The key length can be 128-bit, 192-bit and 256-bit. The number of rounds also varies 10 for a 128-bit key, 12 for a 192-bit key, and 14 for a 256-bit key. There are 4 phases of AES encryption: Substitute Bytes, Shift Row, Mixed Columns and Add Round Key.

AES is one of the most powerful algorithms that are widely used in different fields all over the world. This algorithm enables faster than DES and 3DES algorithms to encrypt and decrypt data. AES is effective in both hardware and software, but it also has a security issue that is Side Channel attacks risk.

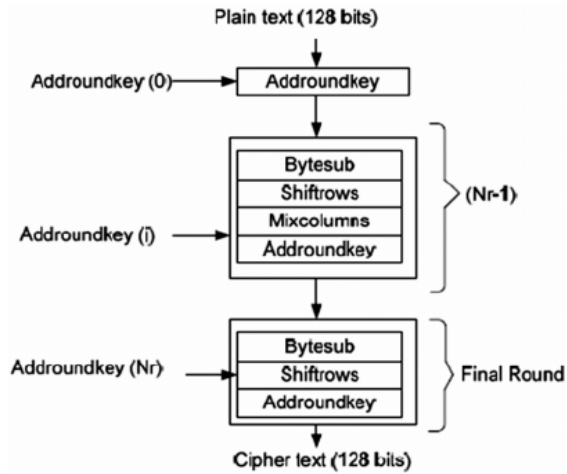

*Fig 4: AES Structure*

**3.4. Blowfish** was developed in 1993 by Bruce Schneier. The block size is equal to 64 bits and the key size varies on a scale from 32-bit to 448-bit. It defines 2 distinct boxes: S boxes, a P box, and four S boxes. Taking into consideration the P Box P is a one-dimensional field with 18 32-bit values. The boxes contain variable values; those can be implemented in the code or generated during each initialization. The S boxes S1, S2, S3, and S4 each contain 256 32-bit values. [9]

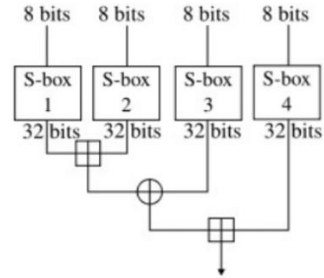

*Fig 5: Blowfish Architecture*

| Advantages | Disadvantages |
|---|---|
| *High speed* | *Needs a secure channel to transmit the secret key* |
| *Simpler* | *Hard to manage different keys* |
| *Requires fewer computer resources* | *Cannot provide digital signatures that cannot be repudiated* |
| *Uses different keys for each pair* | *Attacker can decrypt both received and sent messages* |

### 4. Asymmetric encryption

Asymmetric encryption is the type of encryption where two different keys, public and private, are used to encrypt and decrypt data. Public key is used by encoder to encrypt the information and private key is used by decoder to decrypt it. Both of these keys are generated by receiver.

Most commonly asymmetric algorithms are RSA, DSA, ElGamal

**4.1. RSA** – is the most commonly used asymmetric encryption algorithm. It was invented by Ron Rivest, Adi Shamir, and Len Adleman and published in 1977. RSA is applied where digital data is used, like Web browsers and servers. It is also used in such security protocols as SSL/TLS, PGP, IPSEC, SILK and SSH. In addition, it can be implemented in emails for authentication and privacy, but most importantly, it is used in electronic credit card payments. One of the important features of RSA is Digital Signature. To verify that the message is indeed from the sender and not coming from an unknown user, it should have a signature. This move also assures that the sender will not be able to deny that the message is coming from him.

The stages of RSA are:
1. Generating 2 random prime numbers p and q
2. Calculating $n = p \times q$
3. Calculating $\varphi(n) = (p-1) \times (q-1)$
4. Generating variable e such that $gcd(\varphi(n), e) = 1$
5. Calculate $d = e^{-1} \, mod(\varphi(n))$

As a result, we get *(e, n)* as a public key and *(d)* as a private key.

One of the possible attacks on RSA can be guessing the *d* number and that can lead to loss of confidentiality of the data, but practically the probability of a successful attack is very low.

The other popular attack is cycle attack. The attacker keeps encrypting the cipher text until the original plaintext is generated. It is important to remember the number of iteration to, later on, use it to decrypt any other cipher text from this sender.

***4.2. DSA (Digital Signature Algorithm)*** – is an application of asymmetric encryption applied on digitalized documents.

The purpose of Digital Signature is to identify the sender and prevent him from denying his actions.

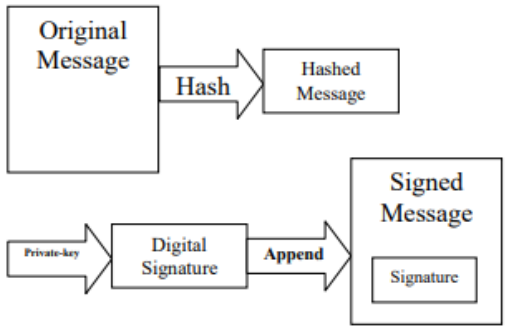

*Fig 6: Steps of generating Digital Signature*

There are several types of Digital Signature Schemes:
1. *Batch Scheme:*
   This scheme generates a random number to verify the identity of the sender. Although theoretically, it is possible to guess this number, practically the probability of doing it is very low.
2. *Forward-Secure Scheme:*
   The security level of this scheme is considered as very high because the secret key is being updated regularly.

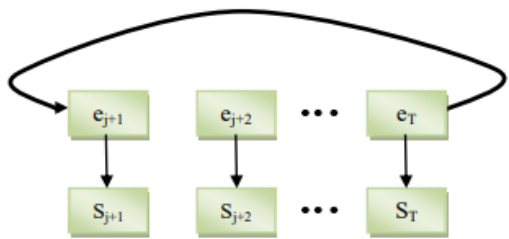

*Fig 7: Forward-Secure Scheme steps*

3. *Blind Scheme:*
   The sender sends a message to the receiver who should sign it but the last one should not know who the sender is.
4. *Proxy Scheme:*
   In case the signer is not able to sign the message, someone he can give the privilege to someone else that he trusts.

Among these schemes, the most secure one is Forward-Secure Scheme, the most difficult one is Proxy Scheme and the most efficient one is Blind Scheme.

**5. Secure connection between patient and doctor**

How does telemedicine work? The patient has all of their biomedical measurements in a database that is located in their home network. As they need to share this data with the doctor, they send it through the internet and know the information is at the remote doctor. To be sure that the data will not be available to a third user it must be encrypted. To decide which encryption type will suit more first we should learn about the structure of this system. After receiving the biomedical measurements, they should be added in a database with the users' identity. To track the change in patients' results these numbers should be updated regularly. As a result, we will get many rows in the table and to encrypt it efficiently it is better to use symmetric encryption. Why? According to the research, we have done before, asymmetric encryption or in other words, public key encryption is slower than symmetric encryption because of its complex algorithms. The system should analyze the information fast to give doctors' recommendation so it should be transferred as soon as possible. In addition, one of the advantages of using symmetric encryption is that the doctor will share only one key with each patient but it also has an issue: if an attacker will have one of the keys, all of the information from and to this patient will be available to them. To guarantee that the user is indeed the patient digital signature can be implemented. The encrypted text will be hashed and sent to the other side and this time we will not have a problem with the amount of data because after encryption the text will be fixed sized.

| Advantages | Disadvantages |
|---|---|
| *Doesn't need a secret channel for the transfer of the public key* | *Needs much money for encryption and decryption of messages* |
| *Creates less key-management problems* | *Low speed* |
| *Can provide electronic signature* | *Uses long keys to provide security* |
| *Uses two different keys that provide communication between sender and receiver, even of their lines of communication are being observed by a third party* | *Cannot provide encryption for large amounts of data* |

## 6.Conclusion:

According to the mentioned information above, we can highlight several differences between symmetric and asymmetric encryption that can help us to decide which encryption type is better to use in some applications or systems, in our case in telemedicine. To reduce the processing time first, we use symmetric algorithms and then to improve the security – asymmetric algorithm such as digital signature.